\documentclass[sigconf]{acmart}
\AtBeginDocument{%
  }

\setcopyright{none}
\renewcommand\footnotetextcopyrightpermission[1]{} 
\begin{document}

\title{Attributes to Support the Formulation of Practically Relevant Research Problems in Software Engineering}

\author{Anrafel Fernandes Pereira}
\affiliation{%
  \institution{PUC-Rio and Univassouras}
  \city{Rio de Janeiro}
  \country{Brazil}}
\email{afpereira@inf.puc-rio.br}

\author{Maria Teresa Baldassarre}
\affiliation{%
  \institution{University of Bari}
  \city{Bari}
  \country{Italy}}
\email{mariateresa.baldassarre@uniba.it}

\author{Daniel Mendez}
\affiliation{%
  \institution{Blekinge Institute of Technology and fortiss}
  \city{Karlskrona}
  \country{Sweden}}
\email{daniel.mendez@bth.se}

\author{Jürgen Börstler}
\affiliation{%
  \institution{Blekinge Institute of Technology}
  \city{Karlskrona}
  \country{Sweden}}
\email{jurgen.borstler@bth.se}

\author{Nauman bin Ali}
\affiliation{%
  \institution{Blekinge Institute of Technology}
  \city{Karlskrona}
  \country{Sweden}}
\email{nauman.ali@bth.se}

\author{Rahul Mohanani}
\affiliation{%
  \institution{M3S, University of Oulu}
  \city{Oulu}
  \country{Finland}}
\email{rahul.mohanani@oulu.fi}

\author{Darja Smite}
\affiliation{%
  \institution{Blekinge Institute of Technology}
  \city{Karlskrona}
  \country{Sweden}}
\email{darja.smite@bth.se}

\author{Stefan Biffl}
\affiliation{%
  \institution{Institute of Information Systems Engineering (TU Wien)}
  \city{Vienna}
  \country{Austria}}
\email{stefan.biffl@tuwien.ac.at}

\author{Rogardt Heldal}
\affiliation{%
  \institution{Western Norway University of Applied Sciences}
  \city{}
  \country{Norway}}
\email{rohe@hvl.no}

\author{Davide Falessi}
\affiliation{%
  \institution{University of Rome "Tor Vergata"}
  \city{Rome}
  \country{Italy}}
\email{d.falessi@gmail.com}

\author{Daniel Graziotin}
\affiliation{%
  \institution{University of Hohenheim}
  \city{Stuttgart}
  \country{Germany}}
\email{graziotin@uni-hohenheim.de}

\author{Marcos Kalinowski}
\affiliation{%
  \institution{PUC-Rio}
  \city{Rio de Janeiro}
  \country{Brazil}}
\email{kalinowski@inf.puc-rio.br}

\renewcommand{\shortauthors}{Pereira \textit{et al.}}

\begin{abstract}
[Background] A well-formulated research problem is essential for achieving practical relevance in Software Engineering (SE), yet there is a lack of structured guidance in this early phase. [Aims] Our goal is to introduce and evaluate seven attributes identified in the SE literature as relevant for formulating research problems – \textit{practical problem}, \textit{context}, \textit{implications/impacts}, \textit{practitioners}, \textit{evidence}, \textit{objective}, and \textit{research questions} – in terms of their perceived importance and completeness, and learn how they can be applied. [Method] We conducted a workshop with 42 senior SE researchers during the ISERN 2024 meeting. The seven attributes were presented using a Problem Vision board filled with a research example. Participants discussed attributes in groups, shared written feedback, and individually completed a survey assessing their importance, completeness, and suggestions for improvement. [Results] The findings confirm the importance of the seven attributes in the formulation of industry-oriented research problems. Qualitative feedback illustrated how they can be applied in practice and revealed suggestions to refine them, such as incorporating financial criteria (\textit{e.g.}, ROI) into implications/impacts and addressing feasibility and constraints under evidence. [Conclusion] The results reaffirm the importance of the seven attributes in supporting a reflective and context-aware problem formulation. Adapting their use to specific research contexts can help to improve the alignment between academic research and industry needs.
\end{abstract}



\maketitle
\section{Introduction}
The relevance of software engineering (SE) research is often undermined by a misalignment between the focus of the research community and the real needs of practitioners \cite{garousi2020}. Studies suggest that improving industrial relevance requires close collaboration between researchers and industry to tackle important real-world problems~\cite{franch2020,garousi2020,winters2024}. However, building such collaboration remains challenging, which contributes to the perception that SE research lacks practical value~\cite{garousi2020}. This issue has been a recurring topic of discussion within the empirical SE community over the past years~\cite{garousi2020,ivarsson2011,molleri2023,petersen2024,winters2024}. 

Garousi \textit{et al.}~\cite{garousi2020} identify that the formulation of research problems that fail to capture the industry needs or make simplistic assumptions about SE in practice are a contributing factor to the lack of practical relevance of SE research. Marijan \& Gotlieb~\cite{marijan2020}  highlight the importance of involving professionals in the problem formulation process so that the solution sought can reflect the complexities of industrial practice. The practical relevance of problems that SE research has focused on is a recurring concern, but previous studies have primarily concentrated on evaluating the impact of research outcomes~\cite{ivarsson2011,molleri2023,petersen2024}, placing less emphasis on the early stage of problem formulation. 

Gorschek \textit{et al.}'s \cite{gorschek2006} present a technology transfer model that offers an approach to connecting academic research to industrial practice. The focus of their model is on guiding co-production, rather than on how to formulate research problems. In this research, we aim to guide research problem formulation to ensure that research studies consider contextual, economic, and human factors to the extent they deserve attention~\cite{ali2016,ali2019,molleri2023,maartensson2016}, helping to foster the alignment of SE research with industrial challenges.

In this paper, we introduce and evaluate seven attributes~--~\textit{practical problem} (what / how / why), \textit{context} (where / when), \textit{implications / impacts} (why), \textit{practitioners} (who), \textit{evidence} (how), \textit{objective} (what / how), and \textit{research questions} (what)~--~to support the formulation of practically relevant research problems in SE, aiming to enrich the problem formulation stage of the Gorschek \textit{et al.}~\cite{gorschek2006} model.

The selection of attributes was based on structured discussions among the first four authors, who reviewed and compared key aspects emphasized in the SE literature related to the formulation of practically relevant research problems~\cite{garousi2020, gorschek2006, ivarsson2011, petersen2024}. Although no formal consensus technique was applied, inclusion decisions were guided by the recurrence and the level of support found in foundational works. The resulting attributes reflect the importance of the definition of the initial pain point or opportunity (\textit{e.g.}, the practical problem)\cite{ gorschek2006,shaw2002}, contextualization\cite{runeson2012case, stol2018, wieringa2014}, practical impact~\cite{lo2015}, practitioners’ identification~\cite{dyba2005}, empirical evidence~\cite{wohlin2013}, well-defined objectives~\cite{basili1988, wieringa2014}, and precise research questions~\cite{basili1988, shaw2002}. Other potential attributes (\textit{e.g.}, method, cost-effective, cost-benefit, scale of the problem, implementable) were considered but excluded at this stage due to conceptual overlap or limited representation in the literature. These attributes were intentionally left open for participants to suggest during the empirical evaluation.

To explore the practical relevance of the proposed attributes, we conducted an empirical study involving 42 senior researchers during the 2024 annual meeting of the International Software Engineering Research Network (ISERN\footnote{https://isern.iese.de/}). This setting was particularly suited for our investigation, as ISERN brings together researchers with extensive experience in industry-academia collaboration. In this paper, we present the conceptual foundation of the seven attributes and describe our preliminary empirical evaluation, which aimed to understand how these attributes are perceived in terms of their importance and completeness when supporting the formulation of practically relevant research problems in SE. The insights derived from this study are intended to guide researchers in making the implicit aspects of problem formulation explicit and actionable.

\section{Related Work}
The outcomes of SE research are commonly intended for industrial application, making it essential to align research with industry needs to ensure practical relevance \cite{winters2024}. To achieve this, research must employ appropriate strategies that balance theoretical advancement with real-world applicability \cite{stol2018}. Previous studies have offered valuable models and frameworks for evaluating practical relevance and improving research communication. 

Ivarsson \textit{et al.} \cite{ivarsson2011}, for example, proposed a model to assess the rigor and relevance of technology evaluations. The model was evaluated through a systematic review of requirements engineering technology evaluations, revealing that most studies lacked rigor and industrial relevance. In addition, they observed that the research field did not show any improvements in terms of industrial relevance over time.

Storey \textit{et al.} \cite{storey2017} proposed a visual abstract template to improve the clarity and dissemination of Design Science research in SE by structuring key elements like problem definition, solution, evaluation, and contributions. Engström \textit{et al.} \cite{engstrom2020} reviewed 38 award-winning SE papers and identified five types of contributions based on design science principles. Both studies highlight that adopting a design science lens improves clarity and novelty, and targets practical relevance as important outcomes.

Petersen \textit{et al.}~\cite{petersen2024} proposed the Reasoning Framework for Relevance (RFR) to improve the design, reporting, and assessment of industrial relevance in software engineering research. Recognizing the lack of consensus on defining and measuring relevance, the authors review different perspectives and key attributes for the concept of relevance, such as applicability, context, and practical impact. The proposal is based on a framework with six aspects (\textit{what, how, where, who, when, and why}), allowing a structured analysis of the industrial research relevance. 

Garousi \textit{et al.} \cite{garousi2020} argue that many studies fail to achieve this alignment due to inadequately formulated research problems. Their multivocal literature review (MLR) identified poor problem definition, weak industry engagement, and simplistic assumptions about practice as key barriers to practical relevance. Recommendations for improving this relevance include using appropriate research approaches, choosing practical problems, and collaborating with industry. Similarly, Winters \cite{winters2024} critiques the disconnect between SE research and industry needs, highlighting issues such as solutions lacking practical context, addressing questions irrelevant to practitioners, and focusing on overly narrow problems. The paper also questions study scalability, arguing that small datasets and niche domains fail to reflect real-world challenges. To bridge this gap, the author advocates for research on broad, high-impact topics like productivity, testing, design, and collaboration—areas that meet the \textit{toothbrush test} by being widely and frequently used. The study calls for a shift toward developing tools, techniques, and processes that provide measurable, practical value to industry.

Although previous studies have made significant contributions to evaluating practical relevance (such as Ivarsson \textit{et al.} \cite{ivarsson2011} and Petersen \textit{et al.} \cite{petersen2024}) and improving research communication (such as Storey \textit{et al.} \cite{storey2017} and Engstrom \textit{et al.} \cite{engstrom2020}), they do not directly address how to support the formulation of practically relevant research problems, a foundational step that shapes the entire research process. Garousi \textit{et al.} \cite{garousi2020} and Winters \cite{winters2024} highlight the importance of this gap by pointing out that poorly defined research problems are among the main reasons for the low applicability of SE research.

\section{Study Design}\label{sec:design}
\begin{figure}[ht]
    \centering
    \includegraphics[width=1\linewidth]{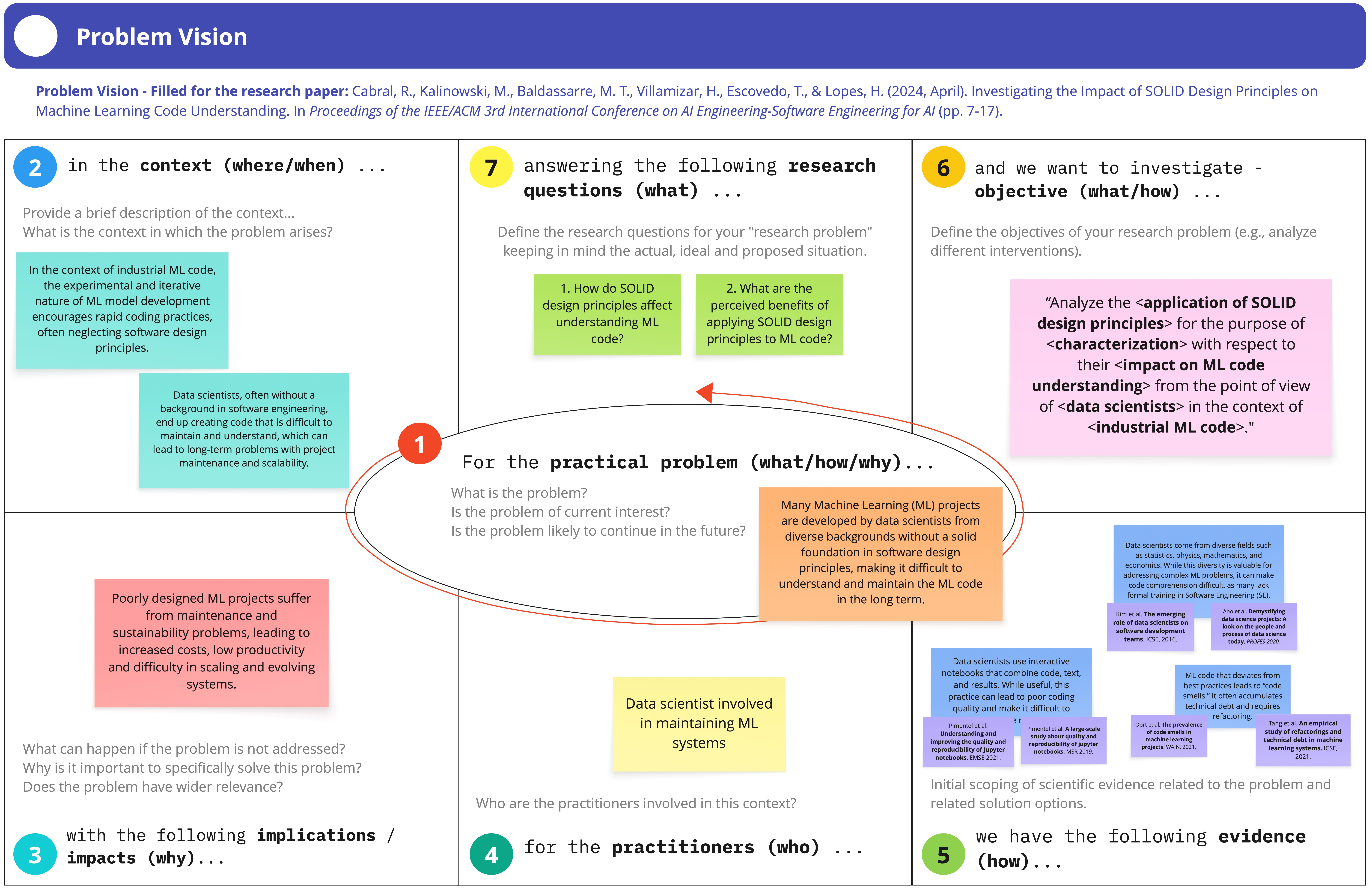}
    \caption{Problem Vision board}
    \label{fig:tpv}
\end{figure}

The seven candidate attributes were organized into a visual board called Problem Vision. This board presents the attributes in an order that supports the structured formulation of the problem. It follows this structure: \textit{``For the practical problem […] involved in the context […] which brings the following implications/impacts […] for the practitioners […], we have the following evidence […]. We want to investigate – objective […] answering the following research questions […]''}. This structure is intended to ensure coherence among the attributes. A visual artifact from Problem Vision supports its practical application (Figure \ref{fig:tpv}) and is available in our open science repository~\cite{fernandes_pereira_2025}. Considering these attributes during the problem formulation stage is important for (1) structuring research problems in a clear and practice-oriented way and (2) ensuring that elements such as target audience, impact, and evidence are made explicit and connected to real-world industry challenges.
Although the study is more accurately characterized as a small-scale evaluation rather than a case study~\cite{wohlin2022case}, given that it was not conducted in a real-life context, we followed the case study design reporting structure recommended by Runeson \textit{et al.}~\cite{runeson2012case}. It should be noted that small-scale evaluations fit well to research where something is proposed, and its (prospective) value needs to be assessed~\cite{wohlin2022case}.

\subsection{Goal and Research Questions}
We defined the goal of this study using the goal definition template of the Goal Question Metric (GQM) paradigm \cite{basili1988} as follows:

``\textbf{Analyze} the seven attributes \textbf{for the purpose of} characterization \textbf{with respect to} the perception of their importance and completeness to support the formulation of practically relevant SE research problems \textbf{from the point of view of} senior SE researchers with experience in industry-academia collaborations \textbf{in the context of} discussing problem formulation based on the attributes for a published research paper and providing feedback on the importance of the attributes."

To investigate this goal, we formulate the following research questions:
\begin{itemize}
    \item \textbf{RQ1:} To what extent are the seven attributes perceived as important to support the formulation of practically relevant SE research problems?
    
    \item \textbf{RQ2:} What other attributes are perceived as important to support the formulation of practically relevant SE research problems?
\end{itemize}

\subsection{Case and Subject Selection}\label{sec:participants}
We designed this study based on the opportunity to engage with 42 senior SE researchers during an in-person workshop held at the 2024 annual meeting of the International Software Engineering Research Network (ISERN). ISERN is a global network of leading researchers in empirical SE who are recognized for their strong commitment to bridging the gap between research and industrial practice. Members of ISERN typically have extensive experience conducting industry-relevant research, collaborating with practitioners, and applying empirical methods in real-world settings. Therefore, the ISERN meeting provided a unique and qualified environment for exploring the importance and completeness of attributes that support the formulation of practically relevant research problems. While no formal demographic survey was conducted, participation in ISERN is by invitation, and membership is contingent upon a track record of active engagement in academia–industry collaboration. This context ensures a high level of familiarity among participants with the challenges of industrial practice, supporting the credibility and relevance of the insights gathered during the study.

In this specific thematic workshop, we present the Problem Vision board. To illustrate its application, we provided participants with a pre-filled example based on a published scientific paper~\cite{cabral2024}. Participants were then invited to reflect on the importance of each attribute to support the problem formulation process and to suggest improvements or additional attributes based on their own experiences. The selected research paper~\cite{cabral2024} addresses a practical and current issue: the maintainability of machine learning (ML) code developed in industry. It investigates how applying software engineering design principles can improve understanding of ML code, which is often written by data scientists with varied backgrounds and limited adherence to design best practices~\cite{staron2024bringing}.

\subsection{Instrumentation}\label{sec:artifacts}
We carefully designed and reviewed all materials for this study to ensure the reliability of data collection. To support this process, we conducted a pilot workshop with twelve master’s and Ph.D. students experienced in Experimental and Empirical SE at ExACTa\footnote{https://www.exacta.inf.puc-rio.br} PUC-Rio. Their familiarity with empirical research enabled them to provide valuable feedback on the clarity, consistency, and usability of the artifacts before their use in the main study.

The study instrumentation consisted of a slide presentation and three main artifacts structured in three envelopes, each corresponding to a specific activity, to guide participants through a step-by-step collaborative process. The first envelope included a pre-filled A2-format Problem Vision board and the article by Cabral \textit{et al.}~\cite{cabral2024}, which participants reviewed and discussed in the first activity, documenting their insights on the board using post-its. The second envelope included a semantic differential scale~\cite{heise1970} to assess the previously discussed research problem based on \textit{value}, \textit{applicability}, and \textit{feasibility}, which participants applied during the second activity. The third envelope included a Likert-type scale survey with optional open-ended questions to evaluate the importance and completeness of attributes and criteria, and to gather suggestions for additional elements not covered by the framework. Participants were divided into five heterogeneous groups based on a mix of institutional affiliation, geographic region, and academic seniority, in order to promote diverse perspectives and balanced discussions. Each group received the three envelopes. All materials are anonymously available in our open science repository~\cite{fernandes_pereira_2025}.

In this paper, we focus on presenting the results obtained from the first and third activities of the workshop, as these phases directly addressed the evaluation of the proposed attributes through group discussion and individual survey responses.

\subsection{Data Collection and Analysis Procedures}
We selected the participants as described in Section~\ref{sec:participants}. During the workshop, they were organized in groups and used the set of artifacts mentioned in Section~\ref{sec:artifacts}. These artifacts facilitated data collection, capturing quantitative and qualitative feedback.

Our evaluation approach combined statistical analysis of Likert-type scale responses with qualitative analysis of open-ended feedback to comprehensively assess participant perceptions. All data were organized in a spreadsheet~\cite{fernandes_pereira_2025} recording responses from the Likert-type scale survey and open-ended survey responses. For quantitative analysis, the survey Likert-type scale agreement options for each attribute were statistically analyzed to determine their frequency. For qualitative analysis, the open-ended survey responses were analyzed to extract information, justify quantitative conclusions, explore nuances in participant perspectives, and identify areas where opinions differed.

This triangulation of quantitative and qualitative data ensured a robust analysis by not only quantifying agreement levels but also contextualizing the reasoning behind participant responses.

\section{Results}
\subsection{Case and Subject Description}
This study focuses on presenting and evaluating the importance of seven attributes~--~\textit{practical problem} (what/how/why), \textit{context} (where/when), \textit{implications/impacts} (why), \textit{practitioners} (who), \textit{evidence} (how), \textit{objective} (what/how), and \textit{research questions} (what)~--~to support formulating research problems relevant to industrial practice. The evaluation was carried out through a 90-minute collaborative workshop at ISERN 2024, structured into five steps: (i) an introductory presentation (20 min); (ii) group formation and material distribution (5 min); (iii) discussion of the research problem using the Problem Vision board (35 min); (iv) evaluation of the relevance of the problem (15 min); and (v) a survey (15 min).

\begin{figure}[ht]
    \centering
    \includegraphics[width=1\linewidth]{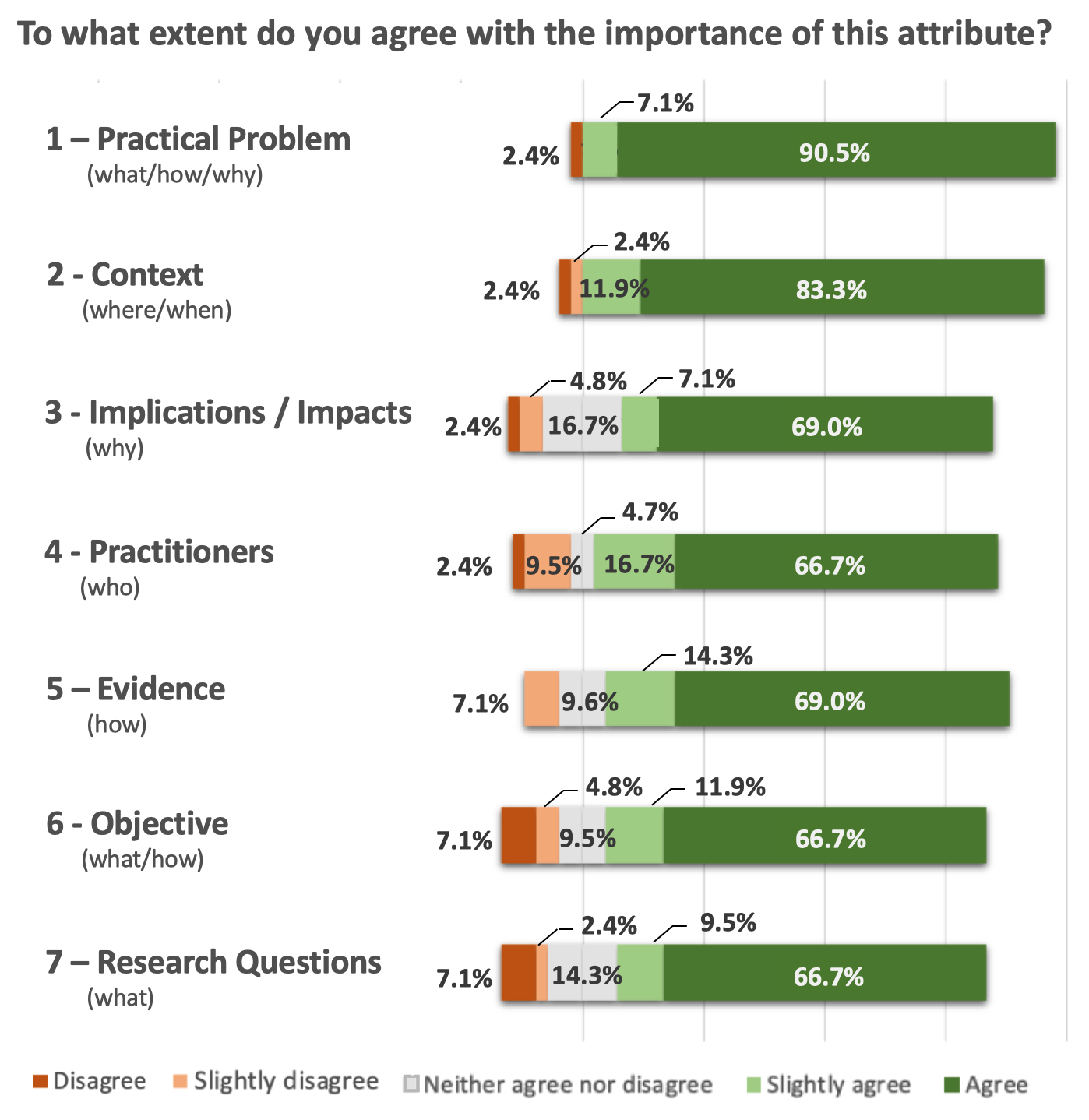}
    \caption{Questionnaire results}
    \label{fig:chart}
\end{figure}

The results are organized around two research questions: RQ1 examines the perceived importance of each attribute, while RQ2 explores additional attributes suggested by participants. The analysis combines quantitative data from the Likert-type scale (see Figure~\ref{fig:chart}) with qualitative insights from open-ended responses, revealing how the attributes were interpreted and applied in the context of industry-academia collaboration.

\subsection{Perceived importance of attributes (RQ1)}

\textbf{Practical Problem (what/how/why):} The \textit{practical problem} attribute received a broad consensus among participants, with 90.5\% selecting ``Agree'' and 7.1\% ``Slightly Agree,'' confirming its central role in formulating relevant research in SE. Qualitative feedback revealed important nuances about how this attribute should be understood and applied in practice.
\begin{itemize}
    \item \textit{``Make sure it is investigated in depth and presented concisely.''} (P10)
    \item \textit{``The problem needs to be a problem from the stakeholder perspective (not [just] from a researcher).''} (P15)
    \item \textit{``Where is the origin of this idea (researcher or practitioners)?''} (P31)
    \item \textit{``Sometimes, the investigations of problems and their implications seem more interesting than the problem itself, and the problem may appear later.''} (P09)
\end{itemize}

Although the importance of the \textit{practical problem} attribute is widely acknowledged, the participants emphasized the need for it to be clearly communicated, grounded in stakeholder perspectives, and sensitive to the origin and context of the idea. Some also suggested that, in certain cases, exploring implications may precede fully defining the problem. These reflections point to the need for a dynamic and iterative approach to problem formulation that aligns both with practical concerns and evolving research contexts.

\textbf{Context (where/when):} The \textit{context} attribute was also broadly recognized as important for formulating practically relevant research problems, with 83. 3\% of the participants selecting ``Agree'' and 11.9\% ``Slightly Agree.'' This reflects a shared understanding that situating a problem in time and space is essential to defining its scope and ensuring applicability. Qualitative comments revealed important insights.
\begin{itemize}
    \item \textit{``Extremely important in SE. Everything is context dependent.''} (P05)
    \item \textit{``Focus on interaction between problem and context.''} (P10)
    \item \textit{``Guidance to define context might help to describe [a] context in an appropriate … and detail level.''} (P33)
    \item \textit{``Helps bound this question, but might be best to not be a focus.''} (P35)
\end{itemize}

Participants emphasized that understanding and articulating context is critical in SE, where solutions often depend on specific situational factors. Some suggested that context should be presented before the problem to improve clarity, while others called for more guidance to help determine the appropriate level of detail. One participant questioned whether context should be a focal point, indicating that its perceived relevance may vary depending on the research scenario. Overall, the findings support the value of this attribute while pointing to areas for refinement in how it is applied.

\textbf{Implications/Impacts (why):} The \textit{implications/impacts} attribute was generally recognized as important, with 69.0\% of participants selecting ``Agree'' and 7.1\% ``Slightly Agree.'' However, this attribute showed more variability than others, with 16.7\% indicating ``Neither agree nor disagree,'' suggesting that the perception of impact is influenced by the nature and context of the research. Participants' comments provided some important insights.
\begin{itemize}
    \item \textit{``Look for real cases where the problem happens. If numbers (money, people, etc.) are provided, they will help to better measure the impact.''} (P06)
    \item \textit{``Any way to quantify the impact, even if it is an estimate? Converting the issue into \$ or person-months.''} (P31)
    \item \textit{``This is why it’s important. I think this will steer the discussion.''} (P35)
    \item \textit{``I am all for basic science research. Not all studies should have practical implications.''} (P12)
    \item \textit{``Hard to predict in a given company.''} (P20)
    \item \textit{``Add ‘Implication from whose perspective?’''} (P29)
\end{itemize}

Participants who agreed with the attribute emphasized the value of quantifying impacts through real-world cases, costs, or effort estimates, reinforcing its role in justifying the relevance of a research problem. Others, especially those neutral or slightly disagreeing, highlighted challenges such as the difficulty of predicting impact in specific contexts, the legitimacy of basic research without immediate practical implications, and the need to clarify the stakeholder perspective. These insights suggest that, while the attribute is essential in practice-oriented research, it requires flexibility, contextualization, and clearer guidance to be applied effectively.

\textbf{Practitioners (who):} The \textit{practitioners} attribute was considered important by the majority of participants, with 66.7\% selecting ``Agree'' and 16.7\% ``Slightly Agree.'' However, it also showed some disagreement, reflecting different views on its applicability across research contexts. The quotations revealed important findings.
\begin{itemize}
    \item \textit{``It’s important to do some profiling and provide some information of characterization.''} (P06)
    \item \textit{``Which role(s) are affected? Developer, tester, manager, business analyst, etc. We have to further define the practitioner’s subcategories.''} (P31)
    \item \textit{``In addition to people involved in the context, perhaps it’s interesting to include people affected by the problem or motivated to solve it.''} (P27)
    \item \textit{``Need to know your audience or who you are pitching the idea to.''} (P35)
    \item \textit{``I would consider the stakeholders rather than only practitioners.''} (P33)
    \item \textit{``Not all studies solve their problems directly. It’s life.''} (P12)
\end{itemize}

Participants who agreed highlighted the value of identifying and characterizing the roles of practitioners affected by or involved in the research problem. Suggestions included defining subcategories of practitioners and considering both direct and indirect stakeholders. However, some participants questioned the centrality of practitioners in all studies, arguing for broader stakeholder inclusion or noting that not all research aims to directly solve practical problems. These perspectives suggest that, while this attribute is useful in industry-oriented research, its application should remain flexible and context-sensitive.

\textbf{Evidence (how):}
The evidence attribute was considered important by most participants, with 69.0\% selecting ``Agree'' and 14.3\% ``Slightly Agree.'' However, 9.6\% responded neutrally and 7.1\% indicated slight disagreement, suggesting mixed perceptions about its role and applicability in problem formulation. Open-ended responses brought valuable perspectives.
\begin{itemize}
    \item \textit{``This is the most important attribute. I would expect an explicit mapping between pieces of evidence and any claim in the implications/impacts.''} (P13)
    \item \textit{``Critically important. Shows that by solving/investigating the current issue, we will actually solve the `real' problem.''} (P35)
    \item \textit{``It seems to be based only on literature. I wonder about more informal evidence, e.g., from the particular company.''} (P39)
    \item \textit{``Practitioners may want to rely on their own experience for evidence (at least evidence supporting motivation).''} (P07)
    \item \textit{``Always compare different groups (e.g., practitioners from different companies, backgrounds, seniority, etc.).''} (P28)
    \item \textit{``Practitioners are looking for solutions, ‘evidence’ sounds a bit academic.''} (P20)
    \item \textit{``Practitioners will not use that but rely on researchers.''} (P17)
\end{itemize}

Participants who agreed highlighted the importance of evidence in validating the practical relevance of a research problem, highlighting the clear links between claims and supporting data. Suggestions included using informal or case-specific evidence in addition to the literature. Others noted that practitioners may prefer experiential knowledge over formal evidence or that the term itself may seem overly academic. These views indicate that, while evidence is essential for rigor and credibility, it should be presented in ways that are accessible, contextualized, and meaningful to both academic and practical audiences.

\textbf{Objective (what/how):} 
The objective attribute was considered important by most participants, with 66.7\% selecting ``Agree'' and 11.9\% ``Slightly Agree.'' However, some disagreement and neutral responses suggest different interpretations about its role and how it should be structured. Some comments offered actionable insights.
\begin{itemize}
    \item \textit{``Objective is the metrics part where you define what to measure so you can answer your research questions.''} (P15)
    \item \textit{``GQM would be applicable here. Objectives should have some form of metric. Success criteria needed to assess the objective.''} (P31)
    \item \textit{``This is a very important step that might imply interactions as there could be several statements to be explored and/or discussed before reaching an agreement.''} (P37)
    \item \textit{``The relationship to the RQs should be clearer and explicit.''} (P13)
    \item \textit{``Add explicitly ‘Apply the GQM template’ to assure a good statement of the objective''} (P29)
    \item \textit{``Depends on evidence. Do not use GQM!!''} (P21)
    \item \textit{``Practitioners might not care about ‘how’.''} (P22)
\end{itemize}

Participants generally agreed that clearly defined objectives are essential for guiding problem formulation and measuring success. Several emphasized the value of linking objectives to research questions and using structured approaches such as GQM. However, some raised concerns about the overreliance on formal templates, the clarity of the current presentation, and the limited relevance of methodological details to practitioners. These insights suggest that, while objectives add clarity and direction, their formulation should remain flexible, explicitly connected to research questions, and adaptable to the needs of both academic and industry audiences.

\textbf{Research Questions (what):}
The \textit{research questions} attribute was considered important by most participants, with 66.7\% selecting ``Agree'' and 9.5\% ``Slightly Agree.'' This reinforces its role in defining the focus and boundaries of the research. The comments revealed varying perspectives on its ideal level of specificity, timing, and relevance, especially in practice-oriented contexts.
\begin{itemize}
    \item \textit{``A starting point. Definitely before ‘Objective.’''} (P11)
    \item \textit{``RQ must be derived from the goal.''} (P16)
    \item \textit{``Checking that potential answers are not simple ‘Yes/No’ and that it does not imply too much extra work for involved people.''} (P25)
    \item \textit{``The term ‘overarching questions’ could be useful in this preliminary scenario.''} (P08)
    \item \textit{``Practitioners might not care about ‘how.’''} (P22)
    \item \textit{``Homework for the academic.''} (P40)
\end{itemize}

Participants emphasized that well-defined research questions should come early in the process, guiding the formulation of objectives and ensuring alignment with research goals. Some cautioned against overly narrow or overly complex questions, while others suggested starting with broader, exploratory questions. Comments also revealed skepticism from a practical point of view. Some participants viewed research questions as primarily academic constructs with limited relevance to practitioners. These findings suggest that, while foundational, research questions should be adaptable in scope and language depending on the research stage and target audience.

Overall, the analysis of the data supports our postulate that the seven attributes are important to support the formulation of practically relevant SE research problems. The study provides a positive response to RQ1 (the median for all attributes reflects agreement on their importance). The results indicate that discussing these attributes facilitates the alignment between academia and industry, helping to ensure practical relevance. Qualitative analysis further enriched the findings with valuable insights.

\subsection{Completeness of attributes (RQ2)}
To identify complementary attributes beyond the seven evaluated, the survey included an open-ended question inviting participants to suggest elements they considered relevant to the formulation of practically relevant research problems. This approach aimed to capture individual insights not covered by the Likert-type scale responses. Of the 42 researchers who participated in the study, 14 suggested additional attributes. 

\textbf{Goals:} The suggestion of ``Goals'' by P36 complements the \textit{objective} attribute by reinforcing the importance of explicitly stating the research goals. As noted: \textit{``It could be interesting to have it stated in a separate section to make it clear.''}

\textbf{Current Relevance:} P40 proposed the attribute ``current relevance,'' questioning whether the problem is valid at present or in the near future: \textit{``Is it valid at this point in time, maybe near future.''} This suggestion refines the \textit{practical problem} attribute by deepening the temporal analysis of relevance, which is already considered in the Problem Vision board (cf. Figure~\ref{fig:tpv}).

\textbf{Challenges/Constraints/Limitations and Feasibility:} The suggestions of ``challenges'' (P16), ``constraints/limitations'' (P09), and ``feasibility'' (P38) complement the \textit{evidence} attribute by emphasizing the need to consider practical limitations of the study. As P09 observed: \textit{``The problem and/or how we solve it is shaped by the constraints/limitations associated with the problem.''} P38 added: \textit{``To be able to study it, the research needs to be feasible.''}

\textbf{Motivation:} The contributions related to ``origin of the problem'' (P25) and ``motivation'' (P28) further elaborate on the \textit{practical problem} attribute by suggesting that the origin and rationale for selecting the problem be made explicit. As highlighted: \textit{``Who found the problem in the first place''} (P25) and \textit{``Why selecting a specific problem/technology?''} (P28), reinforcing the alignment with real needs and practical justification.

\textbf{ROI – Return on Investment:} P35's suggestion of ``ROI'' adds an economic perspective to the \textit{implications/impacts} attribute, as illustrated by: \textit{``Why the company/people spending the time/effort/money.''} This deepens the impact analysis by including financial criteria that are relevant to the industry.

\textbf{Stakeholders:} The suggestions from P29 and P14 regarding ``stakeholders'' complement the \textit{practitioners} attribute by broadening the understanding of who is involved or affected by the problem, including actors such as project managers and decision-makers: \textit{``Explicitly refer to a specific group of people having the problem.''}

\textbf{Method:} P03 suggested the ``method'' attribute, arguing that providing clarity about the methods to be applied can enhance collaboration with industry professionals and increase the applicability of research results: \textit{``Maybe you could add some simple method detail to bring practitioners close to academia.''} While we acknowledge the importance of this perspective, the method is typically defined based on the problem, that is, after the research problem has been formulated.

\textbf{Proposed Solution:} P24 proposed the ``proposed solution'' attribute, emphasizing that in certain cases the research effort is more focused on evaluating a solution rather than identifying the problem itself: \textit{``Some problems (like the one presented) are not problems but solution validation tasks.''} This suggestion indicates that, in solution-oriented studies, it might be helpful to consider aspects of the solution early on, even if that extends beyond the traditional boundaries of problem formulation.

RQ2 did not aim for statistical representativeness, but rather to capture diverse expert perspectives capable of enriching existing attributes or suggesting new directions for future research. The suggestions collected represent a valuable qualitative contribution that offers insights that extend both the depth and scope of the original attribute set. Although several suggestions directly complement the seven attributes evaluated—by refining their clarity, scope, or practical orientation—others appear to be more aligned with later stages of the research process, such as method selection or solution validation. Together, these contributions highlight important considerations such as feasibility, stakeholder diversity, and economic justification, reinforcing the need to adapt attribute use to specific research contexts and goals.

\section{Discussion}
The results of RQ1 confirm the perceived relevance of the seven proposed attributes—\textit{practical problem, context, implications/impacts, practitioners, evidence, objective,} and \textit{research questions}—for formulating practically relevant research problems in SE. Beyond affirming their importance, participants’ qualitative feedback provided valuable nuances drawn from real-world experience. \textit{Practical problem} and \textit{context} stood out as foundational, highlighting that well-defined and context-aware problems improve clarity and impact. However, participants also highlighted the need for a dynamic formulation process, reinforcing that the alignment of academic and industrial perspectives is essential for relevance and collaboration.

Attributes such as \textit{implications/impacts} and \textit{practitioners} showed greater variability in responses, revealing challenges in balancing academic rigor with practical concerns. Participants stressed the need to broaden the definition of stakeholders, incorporate informal evidence, and maintain flexibility in defining \textit{objectives} and \textit{research questions}. RQ2 complemented these findings by suggesting refinements, such as \textit{feasibility}, \textit{ROI}, and broader \textit{stakeholder} involvement, that improve the applicability of the original attributes. Some suggestions, such as the \textit{method} and \textit{proposed solution}, pointed to adjacent stages of the research process, indicating the value of adapting attribute use to specific research contexts and goals.

This study contributes to advancing the discussion on the practical relevance of SE research by focusing on the formulation of research problems, a stage still underexplored in the literature. Previous studies, such as Garousi \textit{et al.}\cite{garousi2020}, Heldal\cite{heldal2024}, and Winters~\cite{winters2024}, have identified challenges such as poorly defined problems, limited scalability, and difficulties in applying findings to practice, but do not propose concrete mechanisms to improve problem formulation. Our work addresses this gap by introducing and empirically evaluating seven attributes that guide researchers in formulating research problems with alignment to industry needs. Experienced SE researchers recognized these attributes as important to support this process. While Ivarsson \textit{et al.}\cite{ivarsson2011} provide a valuable model for assessing rigor and relevance, and Petersen \textit{et al.}\cite{petersen2024} propose a reasoning framework for evaluating relevance, our study operationalizes similar dimensions into actionable attributes for the early stages of research design.

Structured within the Problem Vision board, the attributes help to make implicit assumptions explicit, align stakeholder expectations, and guide problem formulation. Qualitative insights from participants indicate their usefulness, especially in practice-oriented research. However, we acknowledge limitations: the perceived importance of each attribute may vary depending on context and participant background, and an overly rigid application may restrict creativity or oversimplify complex scenarios. Since this study focused on the perceptions of senior SE researchers, future work should involve industry practitioners to assess the practical applicability and generalizability of these attributes in real-world settings.

\section{Limitations and Threats to Validity}
We address the four categories of validity threats described by Wohlin \textit{et al.}~\cite{wohlin2024}:

\textbf{Internal Validity:} Group dynamics and the use of structured materials may have introduced conformity bias or encouraged agreement with the attributes presented. To mitigate these effects, participants were divided into heterogeneous groups based on institutional affiliation, geographic region, and academic seniority. Activities included individual reflection, open-ended questions, and space for suggesting new attributes. Additionally, using a single example (an ML code study applying SOLID principles) may have limited the diversity of participant perspectives. Although the example was chosen for its clarity and practical relevance, we acknowledge that multiple cases could have broadened the range of insights. Future studies should include different domains to validate the attributes across varied scenarios.

\textbf{External Validity:} The study was conducted with senior SE researchers from ISERN, which may limit the generalizability of the results. However, since these researchers are experts in empirical software engineering and have experience with industry-academia collaboration, their contributions are highly relevant. Still, we acknowledge that involving participants from other domains or industrial contexts could have revealed different attributes. Therefore, future research should adopt different methodological designs and directly involve industry practitioners to strengthen the validity of the findings.

\textbf{Construct Validity:} The Likert-type scale may not have fully captured participant perceptions, and interpretations of attributes may have varied. To address these threats, open-ended questions were included to gather additional information on the participants' agreement, and workshop materials were carefully reviewed to maintain consistency and avoid ambiguity. Furthermore, the study and the usage of all its instruments were piloted with master and Ph.D. students before they were used in the study, allowing (minor) adjustments to further enhance clarity.

\textbf{Conclusion Validity:} The sample size (42 participants) and reliance on Likert-type scale responses limit the possibilities of using inferential statistics. We are aware of this limitation. However, to strengthen our conclusions, we triangulated quantitative analysis with qualitative analysis of the open-ended questions to allow a more robust interpretation of the findings.

\section{Conclusions}
This study introduced and evaluated seven attributes to support the formulation of practically relevant research problems in SE. These attributes ~--~\textit{practical problem} (what/how/why), \textit{context} (where/when), \textit{implications/impacts} (why), \textit{practitioners} (who), \textit{evidence} (how), \textit{objective} (what/how), and \textit{research questions} (what)~--~were recognized by experienced SE researchers as important for surfacing implicit assumptions, clarifying stakeholder expectations, and fostering a more reflective and context-aware research problem formulation process. 

Furthermore, the results revealed practical and operational concerns that refine the seven attributes, enhancing their relevance to real-world research. Participants suggested incorporating constraints, feasibility, and ROI to deepen the attributes of \textit{evidence} and \textit{implications/impacts}, while also recommending broader stakeholder involvement to extend \textit{practitioners}. In contexts focused on solution validation, refinements like \textit{method} and \textit{proposed solution} were proposed as useful additions. These refinements help tailor the attributes to specific research scenarios, improving their applicability in practice.

Overall, based on the discussions, we believe that the seven attributes (organized in the Problem Vision board) offer a coherent and adaptable structure to guide early-stage problem formulation. Our findings suggest that researchers could consider these attributes as guidance on aspects to be considered throughout problem formulation, especially when engaging in practice-oriented studies. The Problem Vision board template is available in our open science repository~\cite{fernandes_pereira_2025}.

\bibliographystyle{ACM-Reference-Format}
\bibliography{references}

@inproceedings{ali2016,
  author       = {Nauman Bin Ali},
  title        = {Is effectiveness sufficient to choose an intervention?: Considering
                  resource use in empirical software engineering},
  booktitle    = {Proceedings of the 10th {ACM/IEEE} International Symposium on Empirical
                  Software Engineering and Measurement, {ESEM} 2016, Ciudad Real, Spain,
                  September 8-9, 2016},
  pages        = {54:1--54:6},
  publisher    = {{ACM}},
  year         = {2016},
  url          = {https://doi.org/10.1145/2961111.2962631},
  doi          = {10.1145/2961111.2962631},
  bibsource    = {dblp computer science bibliography, https://dblp.org}
}

@article{garousi2020,
  title={Practical relevance of software engineering research: synthesizing the community’s voice},
  author={Garousi, Vahid and Borg, Markus and Oivo, Markku},
  journal={Empirical Software Engineering},
  volume={25},
  pages={1687--1754},
  year={2020},
  publisher={Springer}
}

@article{franch2020,
  title={How do practitioners perceive the relevance of requirements engineering research?},
  author={Franch, Xavier and Mendez, Daniel and Vogelsang, Andreas and Heldal, Rogardt and Knauss, Eric and Oriol, Marc and Travassos, Guilherme H and Carver, Jeffrey C and Zimmermann, Thomas},
  journal={IEEE Transactions on Software Engineering},
  volume={48},
  number={6},
  pages={1947--1964},
  year={2020},
  publisher={IEEE}
}

@inproceedings{petersen2024,
  title={Revisiting the construct and assessment of industrial relevance in software engineering research},
  author={Petersen, Kai and B{\"o}rstler, J{\"u}rgen and Ali, Nauman Bin and Engstr{\"o}m, Emelie},
  booktitle={Proceedings of the 1st IEEE/ACM International Workshop on Methodological Issues with Empirical Studies in Software Engineering},
  pages={17--20},
  year={2024}
}

@article{molleri2023,
  title={Determining a core view of research quality in empirical software engineering},
  author={Moll{\'e}ri, Jefferson Seide and Mendes, Emilia and Petersen, Kai and Felderer, Michael},
  journal={Computer Standards \& Interfaces},
  volume={84},
  pages={103688},
  year={2023},
  publisher={Elsevier}
}

@article{ivarsson2011,
  title={A method for evaluating rigor and industrial relevance of technology evaluations},
  author={Ivarsson, Martin and Gorschek, Tony},
  journal={Empirical Software Engineering},
  volume={16},
  pages={365--395},
  year={2011},
  publisher={Springer}
}

@article{gorschek2006,
  title={A model for technology transfer in practice},
  author={Gorschek, Tony and Garre, Per and Larsson, Stig and Wohlin, Claes},
  journal={IEEE Software},
  volume={23},
  number={6},
  pages={88--95},
  year={2006},
  publisher={IEEE}
}

@book{wohlin2024,
  author       = {Claes Wohlin and
                  Per Runeson and
                  Martin H{\"{o}}st and
                  Magnus C. Ohlsson and
                  Bj{\"{o}}rn Regnell and
                  Anders Wessl{\'{e}}n},
  title        = {Experimentation in Software Engineering, Second Edition},
  publisher    = {Springer},
  year         = {2024}
}

@article{basili1988,
  title={The {TAME} project: Towards improvement-oriented software environments},
  author={Basili, Victor R and Rombach, H Dieter},
  journal={IEEE Transactions on Software Engineering},
  volume={14},
  number={6},
  pages={758--773},
  year={1988},
  publisher={IEEE}
}

@article{engstrom2020,
  title={How software engineering research aligns with design science: a review},
  author={Engstr{\"o}m, Emelie and Storey, Margaret-Anne and Runeson, Per and H{\"o}st, Martin and Baldassarre, Maria Teresa},
  journal={Empirical Software Engineering},
  volume={25},
  pages={2630--2660},
  year={2020},
  publisher={Springer}
}

@book{wieringa2014,
  title={Design science methodology for information systems and software engineering},
  author={Wieringa, Roel J},
  year={2014},
  publisher={Springer}
}

@article{heise1970,
  title={The semantic differential and attitude research},
  author={Heise, David R},
  journal={Attitude measurement},
  volume={4},
  pages={235--253},
  year={1970},
  publisher={Chicago}
}

@book{runeson2012case,
  title={Case study research in software engineering: Guidelines and examples},
  author={Runeson, Per and Host, Martin and Rainer, Austen and Regnell, Bjorn},
  year={2012},
  publisher={John Wiley \& Sons}
}

@inproceedings{cabral2024,
  title={Investigating the Impact of SOLID Design Principles on Machine Learning Code Understanding},
  author={Cabral, Raphael and Kalinowski, Marcos and Baldassarre, Maria Teresa and Villamizar, Hugo and Escovedo, Tatiana and Lopes, H{\'e}lio},
  booktitle={Proceedings of the IEEE/ACM 3rd International Conference on AI Engineering-Software Engineering for AI},
  pages={7--17},
  year={2024}
}

@article{wohlin2022case,
  title={Is it a case study?—A critical analysis and guidance},
  author={Wohlin, Claes and Rainer, Austen},
  journal={Journal of Systems and Software},
  volume={192},
  pages={111395},
  year={2022},
  publisher={Elsevier}
}

@article{winters2024,
  title={Thoughts on applicability},
  author={Winters, Titus},
  journal={Journal of Systems and Software},
  volume={215},
  pages={112086},
  year={2024},
  publisher={Elsevier}
}

@article{ali2019,
  title={On the search for industry-relevant regression testing research},
  author={Ali, Nauman Bin and Engstr{\"o}m, Emelie and Taromirad, Masoumeh and Mousavi, Mohammad Reza and Minhas, Nasir Mehmood and Helgesson, Daniel and Kunze, Sebastian and Varshosaz, Mahsa},
  journal={Empirical Software Engineering},
  volume={24},
  pages={2020--2055},
  year={2019},
  publisher={Springer}
}

@article{maartensson2016,
  title={Evaluating research: A multidisciplinary approach to assessing research practice and quality},
  author={M{\aa}rtensson, P{\"a}r and Fors, Uno and Wallin, Sven-Bertil and Zander, Udo and Nilsson, Gunnar H},
  journal={Research Policy},
  volume={45},
  number={3},
  pages={593--603},
  year={2016},
  publisher={Elsevier}
}

@article{staron2024bringing,
  title={Bringing Software Engineering Discipline to the Development of AI-Enabled Systems},
  author={Staron, Miroslaw and Abrah{\~a}o, Silvia and Lewis, Grace and Muccini, Henry and Honnenahalli, Chetan},
  journal={IEEE Software},
  volume={41},
  number={5},
  pages={79--82},
  year={2024},
  publisher={IEEE}
}

@article{stol2018,
  title={The ABC of software engineering research},
  author={Stol, Klaas-Jan and Fitzgerald, Brian},
  journal={ACM Transactions on Software Engineering and Methodology (TOSEM)},
  volume={27},
  number={3},
  pages={1--51},
  year={2018},
  publisher={ACM New York, NY, USA}
}

@inproceedings{storey2017,
  title={Using a visual abstract as a lens for communicating and promoting design science research in software engineering},
  author={Storey, Margaret-Anne and Engstrom, Emelie and H{\"o}st, Martin and Runeson, Per and Bjarnason, Elizabeth},
  booktitle={2017 ACM/IEEE International Symposium on Empirical Software Engineering and Measurement (ESEM)},
  pages={181--186},
  year={2017},
  organization={IEEE}
}

@inproceedings{heldal2024,
  title={Is generalisation hindering the adoption of your findings?},
  author={Heldal, Rogardt},
  booktitle={Proceedings of the 18th ACM/IEEE International Symposium on Empirical Software Engineering and Measurement},
  pages={348--358},
  year={2024}
}

@article{shaw2002,
  title={What makes good research in software engineering?},
  author={Shaw, Mary},
  journal={International Journal on Software Tools for Technology Transfer},
  volume={4},
  number={1},
  pages={1--7},
  year={2002},
  publisher={Springer}
}

@inproceedings{lo2015,
  title={How practitioners perceive the relevance of software engineering research},
  author={Lo, David and Nagappan, Nachiappan and Zimmermann, Thomas},
  booktitle={Proceedings of the 2015 10th Joint Meeting on Foundations of Software Engineering},
  pages={415--425},
  year={2015}
}

@article{dyba2005,
  title={Evidence-based software engineering for practitioners},
  author={Dyba, Tore and Kitchenham, Barbara A and Jorgensen, Magne},
  journal={IEEE Software},
  volume={22},
  number={1},
  pages={58--65},
  year={2005},
  publisher={IEEE}
}

@article{wohlin2013,
  title={An evidence profile for software engineering research and practice},
  author={Wohlin, Claes},
  journal={Perspectives on the Future of Software Engineering: Essays in Honor of Dieter Rombach},
  pages={145--157},
  year={2013},
  publisher={Springer}
}

@dataset{fernandes_pereira_2025,
  author       = {Fernandes Pereira, Anrafel and
                  Kalinowski, Marcos and
                  Baldassarre, Maria Teresa and
                  Méndez Fernández, Daniel and
                  Börstler, Jürgen and
                  Ali, Nauman bin and
                  Biffl, Stefan and
                  Falessi, Davide and
                  Graziotin, Daniel and
                  Heldal, rogardt and
                  Mohanani, Rahul and
                  Smite, Darja},
  title        = {Artifacts: Attributes to Support the Formulation
                   of Practically Relevant Research Problems in
                   Software Engineering
                  },
  month        = oct,
  year         = 2025,
  publisher    = {Zenodo},
  doi          = {10.5281/zenodo.17168059},
  url          = {https://doi.org/10.5281/zenodo.17168059},
}

@inproceedings{marijan2020,
  title={Lessons learned on research co-creation: making industry-academia collaboration work},
  author={Marijan, Dusica and Gotlieb, Arnaud},
  booktitle={2020 46th Euromicro Conference on Software Engineering and Advanced Applications (SEAA)},
  pages={272--275},
  year={2020},
  organization={IEEE}
}

@String{Computer = "{IEEE} Computer" }

@String{Springer = "Springer-Verlag" }

\appendix
\end{document}